\documentclass[twocolumn,showpacs,amsmath,amssymb,prb,citeautoscript]{revtex4}

\usepackage{graphicx}
\usepackage{times}

\begin{document}

\title{On the Sign Problem in the Hirsch-Fye Algorithm for Impurity Problems}

\author{Jaebeom Yoo}
\author{Shailesh Chandrasekharan}
\author{Ribhu K.~Kaul}
\author{Denis Ullmo}
\thanks{
Permanent address: Laboratoire de Physique Th\'eorique et
  Mod\`eles Statistiques
(LPTMS), 91405 Orsay Cedex, France}
\author{Harold U.~Baranger}
\affiliation{Department of Physics, Duke University, Durham NC 27708-0305}

\date{\today}

\begin{abstract}
We show that there is no fermion sign problem in the Hirsch and Fye algorithm for the single-impurity Anderson model. Beyond the particle-hole symmetric case for which a simple proof exists, this has been known only empirically. Here we prove the nonexistence of a sign problem for the general case by showing that each spin trace for a given Ising configuration is separately positive. We further use this insight to analyze under what conditions orbitally degenerate Anderson models or the two-impurity Anderson model develop a sign.  
\end{abstract}


\maketitle

Quantum impurity problems describe the interaction of an electron gas with an impurity possessing internal degrees of freedom and, therefore, a dynamics of its own \cite{Hew}.  To study the non-trivial physics involved, it has proven useful to develop numerical methods which are exact.  One widely used ``impurity solver'' is the Monte Carlo algorithm developed by Hirsch and Fye for the Anderson model \cite{HirFye,GubHirFye}.  This algorithm is both simple and useful as one only keeps track of the impurity-impurity Green function during the Monte Carlo updates.  Although it was originally developed to study the simplest form of the Anderson Hamiltonian, this algorithm has found application in the context of dynamical mean field calculations \cite{GeoKot} and, more recently, the mesoscopic Kondo problem \cite{Kaul05}.

One reason for the success of the Hirsch-Fye quantum Monte Carlo algorithm is that it is free of problems caused by the negative sign that results upon permuting two fermions. Such problems tend to plague fermionic quantum Monte Carlo algorithms \cite{Troyer_Wiese05,Evertz03}. In the Hirsch-Fye case, the absence of such a ``sign problem'' can be proved easily if particle-hole symmetry is preserved \cite{Hir85,FyeHir88} and has been known empirically in the general case \cite{GeoKot,HirPrivate}. Here we provide a formal proof of the absence of any sign problem in the general case. The proof relies on the simple fact that the single impurity problem is effectively one-dimensional, and so there is no fermionic permutation sign.

The Anderson single-impurity Hamiltonian \cite{And} is
\begin{eqnarray}
  H & = & \sum_{\sigma = \uparrow,\downarrow} H_{0\sigma} + H_1 \;, \\
  H_{0\sigma} &=& \sum_{ k=1}^N \epsilon_{ k \sigma}
          c^\dagger_{ k\sigma} c_{ k\sigma}
	  + \sum_{ k=1}^N \big ( V_{ k}^* 
          c^\dagger_{ k\sigma} d_\sigma  
        +  V_{ k} d^\dagger_\sigma c_{ k\sigma} \big )  \nonumber \\
        & &+\;
         \big(\epsilon_{d\sigma}+\frac{U}{4})
            d_{\sigma}^\dagger d_{\sigma} \; , \\ 
        H_1&=& U\big[n_{d\uparrow}n_{d\downarrow}-\frac{1}{2}
	  (n_{d\uparrow}+n_{d\downarrow})\big] \,.\label{eq:H}
\end{eqnarray}
We do not make any particular assumption on the values of the energies $\epsilon_{k\sigma}$, the couplings $V_k$, or the impurity parameters $U$ and $\epsilon_{d\sigma}$.  Using the Trotter product formula \cite{Tro}, one can calculate the partition function in the limit
\begin{equation}
  Z= \mathrm{Tr} e^{-\beta H} = 
  \lim_{M\rightarrow\infty} \mathrm{Tr}\,\Big( \prod_{\mu=1}^M e^{-\tau H_0}e^{-\tau H_1}\Big)
\end{equation}
where $\tau \!=\! \beta/M$.  For each time slice $\mu  \!=\!  1, \ldots M$, an auxiliary Ising variable $s_\mu$ is introduced to decouple the spin-up and spin-down parts in $H_1$ using the discrete Hubbard-Stratonovich transformation\cite{Hir},
\begin{equation}
     e^{-\tau H_1} =\frac{1}{2}\sum_{s_\mu = \pm 1} 
     e^{\lambda s_\mu (n_{d\uparrow} - n_{d\downarrow})}\,,
\end{equation}
where $\lambda$ is a constant satisfying the relation $\cosh(\lambda) \!=\! \exp(\tau U/2)$.  Then, the partition function can be computed as
\begin{equation}
     Z=\sum_{\{{\bf s}\}} Z_\uparrow({\bf s})Z_\downarrow({\bf s})\,,
\end{equation}
where the sum is over all possible spin configurations ${\bf s}  \!=\!  (s_1,\ldots,s_M)$ and the spin-up and spin-down partition functions, $Z_\uparrow(\bf{s})$ and $Z_\downarrow({\bf s})$, are given by
\begin{equation} \label{eq:Z}
     Z_\sigma({\bf s}) = \mathrm{Tr} \prod_{\mu=1}^M e^{-\tau H_{0\sigma}}
     e^{\theta(\sigma) s_\mu \lambda n_{d\sigma}}
\end{equation}
with $\theta(\uparrow)  \!=\!  1$ and $\theta(\downarrow)  \!=\!  -1$.  In the Hirsch-Fye algorithm, Ising configurations $\{{\bf s}\}$ are generated according to the weight $Z_\uparrow({\bf s})Z_\downarrow({\bf s})$ using a Metropolis accept/reject step.  For the half-filled Anderson model with particle-hole symmetry, it can be shown easily that the product $Z_\uparrow({\bf s})Z_\downarrow({\bf s})$ is always positive \cite{FyeHir88}.  It is, however, empirically found that $Z_\sigma({\bf s})$ itself is always positive for every ${\bf s}$, even in the general case.

To prove this, it suffices to find a basis of the fermionic Hilbert space, independent of the Ising variable $s_\mu$, in which all the matrix elements of $ e^{-\tau H_{0\sigma}} e^{\theta(\sigma) s_\mu \lambda n_{d\sigma}}$ are positive. To do this, we cast the problem in a one-dimensional form, as follows. First, note that the impurity is coupled locally to the electron gas, through the single-particle operator
\begin{equation}
f_{1\sigma} = \frac{1}{\widetilde{V}} \sum_{ k} V_{ k} c_{ k\sigma}
\end{equation}
where $\widetilde{V} \!=\! \sqrt{\sum_{ k} |V_{ k}|^2}$. 
One can therefore  convert $H_0$ to a tridiagonal form of single-particle operators, or equivalently into an open fermion chain \cite{Wil,KriWil,Hew,YooCha},
\begin{equation}
    H_{0\sigma} = - \sum_{j=0}^{N-1} h_{j\sigma}+\Lambda \hat{N}_\sigma,
\end{equation}
where
\begin{equation} \label{eq:h}
h_{j\sigma} = \alpha_j f_{j\sigma}^\dagger f_{j\sigma}
+  \beta_j^* f_{j\sigma}^\dagger f_{j+1\sigma}
+  \beta_j f_{j+1\sigma}^\dagger f_{j\sigma} 
\end{equation}
with $\beta_N \!=\! 0$, $f_{0\sigma} \!=\! d_{\sigma}$, 
and $\beta_0 \!=\! \widetilde{V}$.  The actual values of the parameters $\alpha_j$ and $\beta_j$ for given $\epsilon_k$'s and $V_k$'s can be obtained in practice using the Lanczos algorithm with $f_1$ as the initial operator.  $\hat{N}_\sigma \!=\! \sum_{i} f_{i\sigma}^\dagger f_{i\sigma}$ is the total number operator.

We first note that the  parameter $\Lambda$ can be chosen such that all the $\alpha_j$'s in the spin chain are positive. Furthermore, by writing $\beta_j  \!=\!  |\beta_j|e^{\phi_j}$ and performing the gauge transformation
\begin{equation}
  f_j = \exp\Big( + i \sum_{m < j} \phi_m \Big) \; f'_j 
  \; , \qquad (j \!
  \ge \! 1) \;  ,
\end{equation}
Eq.~(\ref{eq:h}) is cast in a form where the $f_j$'s are replaced by $f'_j$'s and all the coefficients $\beta_j$ are replaced by the positive real numbers $|\beta_j|$. As a consequence, with the occupation number states $\{|n_{0\sigma}, n_{1\sigma}, n_{2\sigma}, \ldots \rangle  \}$ as the basis of the spin-$\sigma$ Fock space (where $n_{j\sigma} \!=\! 0,1$) and the phase convention
\begin{equation}
|n_{0\sigma} \ldots  n_{N\sigma} \rangle  = 
{f'}_{N\sigma}^{\dagger \; n_N} \ldots
 {f'}_{1\sigma}^{\dagger \; n_1} {f'}_{0\sigma}^{\dagger \;n_0} \;
 |O\rangle \;,
\end{equation}
all matrix elements of $-\tau( H_{0\sigma}-\Lambda \hat{N}_{\sigma})$ are positive.
Since the operator $\hat{N}_\sigma$ commutes with $H_{0\sigma}$, all matrix elements of $\exp(-\tau H_{0\sigma}) \!\equiv\! \exp(-\tau [H_{0\sigma} \!-\! \Lambda \hat{N}_\sigma]) \exp(-\tau \Lambda \hat{N}_\sigma)$ are positive, as are those of $\exp[\theta(\sigma) s_\mu \lambda n_{d\sigma}]$. Note that the basis states are eigenstates of $n_{d\sigma}  \!=\!  d^\dagger_\sigma d_\sigma  \!=\!  f^\dagger_{0\sigma} f_{0\sigma}$.  The trace in Eq.~(\ref{eq:Z}) is, then, positive.

Thus because of its essentially one-dimensional nature, the single impurity Anderson model has no sign problem in the Hirsch-Fye Monte-Carlo algorithm for any choice of parameters (in fact each spin determinant is separately positive). These considerations may be applied to a large class of models, thus indicating without any numerical effort whether a model has no sign problem or whether it is likely to have one. We now discuss an example of each case.

(1) \textit{Orbital- and Band- degenerate Anderson model:} Consider an atomic impurity in a metal whose Fermi surface is rotationally invariant (this approximation may be relaxed). Additionally, as is the case for most transition metal or rare earth impurities, let the impurity's outer shell be orbitally degenerate ($l \!=\! 2,3$ as the case may be). Consider a general density-density interaction between the orbitals of the impurity, $\sum_{\alpha \beta} U_{\alpha \beta}n_\alpha n_\beta$ where $\alpha, \beta$ are the combined orbital and spin indices $m\sigma$. This type of orbitally degenerate impurity model was studied, for instance, in Ref.\,\onlinecite{NozBla} (note that $l \!=\! 0$ corresponds to the usual single impurity Anderson model). 

Because of the rotational invariance, for each orbital $\alpha$ on the impurity there is one particular linear combination of electron gas states, $\Psi_{\alpha}$, which couples to it. In fact, $\alpha$ is a conserved quantity, and so we may separate the $\alpha$ channels into independent one-dimensional (1-D) quantum channels, as in Eq.~(\ref{eq:h}), each written as a semi-infinite spin chain. Thus we have $2(2l \!+\! 1)$ channels that are coupled only by the the quartic interaction terms $U_{\alpha\beta}$ at the impurity. Each quartic term may be split into quadratic terms using a discrete Hubbard-Stratonovich \cite{HirFye} field ${\bf s}_{\alpha\beta}$. Then, $Z=\prod_{\alpha=1}^{2(2l+1)} Z_\alpha$ with
\begin{equation}
Z_\alpha({\bf s}_{\alpha \beta}) = \mathrm{Tr} \prod_{\mu=1}^M e^{-\tau H_{0\alpha}} e^{\sum_{\beta}\theta_{\alpha \beta} s^{\mu}_{\alpha\beta} \lambda_{\alpha\beta} n_{\alpha}}
\end{equation}
where $\theta_{\alpha\beta} \!=\! 1$ if $\alpha \!>\! \beta$ and $\theta_{\alpha\beta} \!=\! -\theta_{\beta\alpha}$. Using the same arguments as for the single impurity problem, we can write each $Z_\alpha$ as a product of matrix elements that are positive. Thus we have shown that the orbitally degenerate Anderson model considered here has no sign problem.

(2) \textit{Two-impurity Anderson model:} As an example where the above ideas indicate the presence of a sign problem, consider two Anderson impurities embedded in an electron gas. Since the single impurity problem does not suffer from a sign problem, one may be tempted to naively hope that this is also the case for the two-impurity model. However, since there are two impurities at different locations in the electron bath, the bath may \textit{not} be written as a sum of independent 1-D quantum channels. Thus fermion world lines in imaginary time may permute, leading in general to a sign problem in fermionic quantum Monte-Carlo. In the special case of particle-hole symmetry, it has been shown that the contributions from $Z_\uparrow$ and $Z_\downarrow$ always have the same sign \cite{HirFye,FyeHir89}, and thus there is no sign problem. 

In summary, we have shown that the Hirsch-Fye algorithm for the single impurity Anderson model does not suffer from a sign problem for any value of the temporal discretization $\tau$ or any values of the parameters in the Hamiltonian. The key property underlying the proof is that the local interaction allows one to transform the Hamiltonian into a one-dimensional chain for which there is no fermionic permutation sign. We also considered two more complicated models to show how these ideas can be used either to prove the absence of a sign problem or to explain the reason for the presence of one.

This work was supported in part by the NSF (DMR-0103003).

\end{document}